\documentclass[12pt, preprint]{aastex}
\usepackage{natbib}
\usepackage{rotating, bm}
\usepackage{epsfig}    
\usepackage{amsmath}   
\usepackage{amssymb}   
\usepackage{latexsym}  
\usepackage{dcolumn}
\newcolumntype{.}{D{.}{.}{-1}}

\begin{document}

\shorttitle{The Hanle Effect in Surface Dynamo Simulations}

\title{Determining the Magnetization of the Quiet Sun Photosphere from the Hanle Effect 
and Surface Dynamo Simulations\footnote{\bf To appear in ApJ Letters, Vol. 731, L21, (2011).}}

\shorttitle{The Magnetization of the Quiet Solar Photosphere}

\author{Nataliya Shchukina\altaffilmark{1} and Javier Trujillo Bueno\altaffilmark{2,3,4}}
\altaffiltext{1}{Main Astronomical Observatory, National Academy of Sciences, 27 Zabolotnogo Street, Kiev 03680, Ukraine}
\altaffiltext{2}{Instituto de Astrof\'{\i}sica de Canarias, E-38200 La Laguna,
Tenerife, Spain}
\altaffiltext{3}{Departamento de Astrof\'\i sica, Universidad de La Laguna, Tenerife, Spain}
\altaffiltext{4}{Consejo Superior de Investigaciones Cient\'{\i}ficas (Spain)}\email{shchukin@mao.kiev.ua; jtb@iac.es}

\begin{abstract}
The bulk of the quiet solar photosphere is thought to be significantly magnetized, due to the ubiquitous presence of a tangled magnetic field at subresolution scales with an average strength ${\langle B \rangle}{\sim}100$ G. This conclusion was reached through detailed three-dimensional (3D) radiative transfer modeling of the Hanle effect in the Sr {\sc i} 4607 \AA\ line, using the microturbulent field approximation and assuming that the shape of the probability density function of the magnetic field strength is exponential. Here we relax both approximations by modeling the observed scattering polarization in terms of the Hanle effect produced by the magnetic field of a 3D photospheric model resulting from a (state-of-the-art) magneto-convection simulation with surface dynamo action. We show that the scattering polarization amplitudes observed in the Sr {\sc i} 4607 \AA\ line can be explained only after enhancing the magnetic strength of the photospheric model by a sizable scaling factor, $F{\approx}10$, which implies ${\langle B \rangle}{\approx}130$ G in the upper photosphere. We argue also that in order to explain both the Hanle depolarization of the Sr {\sc i} 4607 \AA\ line and the Zeeman signals observed in Fe {\sc i} lines we need to introduce a height-dependent scaling factor, such that the ensuing ${\langle B \rangle}{\approx}160$ G in the low photosphere and ${\langle B \rangle}{\approx}130$ G in the upper photosphere. 
\end{abstract}


\section{Introduction}

The importance of the Hanle effect as a diagnostic tool of the quiet Sun magnetism 
cannot be overestimated. This magnetic-field-induced 
modification of the linear polarization due to scattering in spectral lines is even sensitive to the presence of magnetic fields that are randomly oriented at scales too small to be resolved (Stenflo 1982). Applying novel diagnostic techniques based on the Hanle effect, Trujillo Bueno et al. (2004) reached the following conclusions (see also Trujillo Bueno et al. 2006):

\begin{itemize}

\item The quiet solar photosphere is permeated by a small-scale tangled magnetic field, whose average strength at heights $h{\approx}300$ km above the visible solar surface is 
${\langle B \rangle}{\approx}130$ G when no distinction is made between granular and intergranular regions.

\item The downward-moving intergranular lane plasma 
is pervaded by relatively strong tangled magnetic fields at subresolution scales, 
with $\langle B \rangle\,{>}\,200$ G.
This conclusion suggests that most of the flux and magnetic energy reside on still unresolved scales.

\item In the upper photosphere the ensuing energy flux estimated using the typical value of $1\,{\rm km}\,{\rm s^{-1}}$ for the convective velocity (thinking in rising magnetic loops) or the Alfv\'en speed (thinking in Alfv\'en waves generated by magnetic reconnection) turns out to be substantially larger than that required to balance the radiative energy losses from the quiet chromosphere. 

\end{itemize}

The first conclusion was obtained by contrasting low resolution observations of the scattering polarization signals of the Sr {\sc i} 4607 \AA\ line with three-dimensional (3D) radiative transfer calculations in a realistic 3D model of the quiet solar photosphere obtained from the hydrodynamical (HD) simulations of Asplund et al. (2000) (hereafter the HD model). The second conclusion resulted from combining the information provided by the strontium line with that we obtained through the application of the Hanle effect line ratio technique for C$_2$ lines explained carefully in Trujillo Bueno et al. (2006). It is however important to note that the above-mentioned HD model is unmagnetized and that, therefore, such conclusions are based on the following hypotheses for the quiet Sun magnetic field that produces Hanle depolarization: (1) the magnetic field is tangled at scales smaller than the mean free path of the line-center photons, with an isotropic distribution of directions and (2) the shape of the probability density function (PDF), describing the fraction of the quiet Sun photosphere occupied by magnetic fields of strength $B$, is exponential (${\rm PDF}(B)={1\over{{\langle B \rangle}}}{{\rm e}^{-B/{\langle B \rangle}}}$). 

The main aim of this paper is to determine the strength of the magnetization of the quiet solar photosphere without making use of such two approximations. To this end, we investigate the Hanle effect of the Sr {\sc i} 4607 \AA\ line in a 3D photospheric model resulting from the magneto-hydrodynamic (MHD) simulations of V\"ogler \& Sch\"ussler (2007), which show a complex small-scale magnetic field that results from dynamo amplification of a weak seed field. These authors went a step further than Cattaneo (1999) by demonstrating that a realistic flow topology of stratified, compressible, surface convection without enforced recirculation is capable of local dynamo action near the solar surface. It is interesting to note that the small-scale ``hidden magnetic field'' inferred via the Hanle effect lends support to the notion of a turbulent solar surface dynamo playing a significant role in the magnetism of the quiet Sun (e.g., Pietarila Graham et al. 2009; see also the reviews by Lites 2009, and by S\'anchez Almeida \& Mart\'\i nez Gonz\'alez 2011). 

\section{Formulation of the Hanle-effect problem in 3D magneto-convection models}

We calculated the number density of Sr {\sc i} atoms at each grid point of a 3D magneto-hydrodynamic model of the solar photosphere resulting from a non-gray radiative transfer version of the Run-C simulation of V\"ogler \& Sch\"ussler (2007) (hereafter, the MHD model). To this end, we solved the standard 3D non-LTE problem without polarization  
including all the allowed radiative and collisional transitions between the 15 bound levels of a realistic model of Sr {\sc i} and the ensuing ionizing transitions to the ground level of Sr {\sc ii}. With the resulting number density of Sr {\sc i} atoms fixed, we solved the full 3D radiative transfer problem of the scattering polarization in the Sr {\sc i} 4607 \AA\ line by applying the complete frequency redistribution theory of resonance line polarization (e.g., Landi Degl'Innocenti \& Landolfi 2004), which is suitable for modeling the fractional linear polarization observed in weak resonance lines like that of Sr {\sc i} at 4607\AA\ (see Sampoorna et al. 2010; and references therein).

The Sr {\sc i} line at 4607 \AA\ results from a single ${^1{\rm S}_0}-{^1{\rm P}_1}$ resonance line transition with $J_l=0$ and $J_u=1$ (the total angular momentum of its lower and upper levels, respectively). Although the ensuing Einstein coefficient for spontaneous emission is very large (i.e., $A_{ul}{\approx}2{\times}10^8\,{\rm s}^{-1}$) the line originates in the solar photosphere because the abundance of strontium is relatively low (see Figure 1). Scattering polarization in the Sr {\sc i} 4607 \AA\ line is caused only by the atomic polarization (population imbalances and quantum coherences) of its upper level, which results from anisotropic radiation pumping in the solar photosphere. The lower level cannot be polarized and the relevant transfer equations for modeling the available low resolution observations of the fractional linear polarization signals are ${{d}\over{d{\tau}}}{X}={X}-S_X$, with $X$ the Stokes parameter $I$, $Q$ or $U$ at the frequency and direction of propagation under consideration  
($d{\tau}=-{\eta_I}ds$ defines the monochromatic optical path along the ray, $\eta_I$ being the absorption coefficient and $s$ the geometrical distance). 

The source function components are $S_I=r\,S^{\rm line}_I+(1-r)B_{\nu}$, $S_Q=r\,S^{\rm line}_Q$ and $S_U=r\,S^{\rm line}_U$, where $B_{\nu}$ is the Planck function and 
$r=\kappa_l\phi_{\nu}/(\kappa_l\phi_{\nu}+\kappa_c)$ (with $\kappa_l$ the line-integrated opacity, $\kappa_c$ the continuum opacity and $\phi_{\nu}$ the normalized absorption profile). The expressions of the line source function components are identical to those used by Trujillo Bueno \& Shchukina (2007), which are functions of the inclination and azimuth of the ray under consideration and of the following position-dependent quantities (with $Q=0$ for $K=0$ and $Q=0,1,2$ for $K=2$): ${S_Q^K}={\frac{2h{\nu}^3}{c^2}}{\frac{2{J}_l+1}{\sqrt{2{J}_u+1}}}{\rho}_Q^K$ (with ${\rho}^K_Q$ the multipolar components of the upper-level density matrix normalized to the overall population of the ground level). Note that ${S_Q^2}$ (with $Q=1,2$) are complex quantities; thus, we will use 
${\tilde{S}^2_Q}$ and ${\hat{S}^2_Q}$ to denote, respectively, the real and imaginary parts of ${S_Q^2}$. The values of such quantities at each spatial grid point have to be found by solving the following coupled set of equations (Manso Sainz \& Trujillo Bueno 1999):

\begin{equation}
S^0_0\,=\,(1-\epsilon){\bar{J}}_0^0\,+\,{\epsilon}\,B_{\nu},
\end{equation}

\begin{eqnarray}
  \left( \begin{array}{l} 
      {S^2_0} \\ 
      {\tilde{S}^2_1} \\
      {\hat{S}^2_1} \\ 
      {\tilde{S}^2_2} \\ 
      {\hat{S}^2_2}
    \end{array} \right) =  {{(1-\epsilon)}\over{1+\delta^{(2)}(1-\epsilon)}} \, w^{(2)}_{J_uJ_l} \, 
  \left( \begin{array}{r}
      {\bar J}^2_0 \\
      \tilde{J}{}^{2}_{1} \\
      -\hat{J}{}^{2}_{1} \\
      \tilde{J}{}^{2}_{2} \\
      -\hat{J}{}^{2}_{2}
    \end{array} \right)\,
    \end{eqnarray}
\vspace{-0.2in}
\begin{eqnarray}
\hspace{0.1in}
- \,{{(1-\epsilon)}\over{1+\delta^{(2)}(1-\epsilon)}}{\Gamma_u}
    \left( \begin{array}{cccccc}
        0 & M_{12} & M_{13} & 0 & 0 \\ 
        M_{21} & 0 & M_{23} & M_{24} & M_{25} \\
        M_{31} & M_{32} & 0 & M_{34} & M_{35} \\
        0 & M_{42} & M_{43} & 0 & M_{45} \\
        0 & M_{52} & M_{53} & M_{54} & 0 
      \end{array} \right) 
    \left( \begin{array}{l}
        {S^2_0} \\ 
        {\tilde{S}^2_1} \\
        {\hat{S}^2_1} \\ 
        {\tilde{S}^2_2} \\ 
        {\hat{S}^2_2}
      \end{array} \right) \nonumber 
\end{eqnarray}
where $\epsilon=C_{ul}/(A_{ul}+C_{ul})$ (with $C_{ul}$ the rate of inelastic collisions with electrons, in ${\rm s}^{-1}$), ${\Gamma_u}=8.79\times10^6g_{u}B/A_{ul}$ (with $g_{u}$ the Land\'e factor of the upper level and 
$B$ the magnetic strength in gauss), $\delta^{(2)}=D^{(2)}/A_{ul}$ (with $D^{(2)}$ the upper-level depolarizing rate due to elastic collisions with neutral hydrogen atoms), and $w^{(2)}_{10}=1$. In these expressions, ${\bar{J}}_Q^K$ (with $Q=0$ for $K=0$ and $Q=0,1,2$ for $K=2$) are the spherical components of the radiation field tensor (Landi Degl'Innocenti \& Landolfi 2004), which are integrals over the frequency and direction of the Stokes parameters weighted by the normalized absorption profile. Thus, ${\bar{J}}_0^0$ quantifies the familiar mean intensity of the incident radiation, ${\bar{J}}_0^2$ its anisotropy, while $\tilde{J}{}^{2}_{Q}$ and $\hat{J}{}^{2}_{Q}$ (with $Q=1,2$) denote, respectively, the real and imaginary parts of ${J}{}^{2}_{Q}$ (which measure the breaking of the axial symmetry of the incident radiation field). The $M_{ij}$-coefficients depend on the inclination ($\theta_B$) of the local magnetic field vector with respect to the vertical Z-axis and on its azimuth ($\chi_B$). 

The depolarizing rate, $\delta^{(2)}=D^{(2)}/A_{ul}$, is proportional to the neutral hydrogen number density (see Eq. 33 in Faurobert-Scholl et al. 1995). Such formula for calculating $\delta^{(2)}$ results from an accurate quantum-mechanical derivation, and we point out that it leads to values similar to those we have obtained through the semi-classical theory of Anstee \& O'Mara (1995). The excellent agreement between the $\delta^{(2)}$ values obtained from the quantum and semi-classical approaches is a clear indication of the reliability of the $\delta^{(2)}$ values of our modeling, because the Sr {\sc i} 4607 \AA\ line is a triplet type transition with spin S=0 (i.e., it is a spectral line for which the classical description holds).  

Of particular interest is the fact that the magnetic strength needed to reach any given depolarization is the larger the greater $\delta^{(2)}$. This can be seen in the expression of the critical magnetic field strength for the onset of the Hanle effect, which can be defined as the strength ($B_c$) for which the Hanle depolarization reduces the zero-field scattering polarization amplitude by about a factor of two. Eq. (A16) of Trujillo Bueno \& Manso Sainz (1999) implies that

\begin{equation}
B_c\,{\approx}\,(1+\delta^{(2)})\,B_H,
\end{equation}
where $B_H=1.137{\times}10^{-7}\,A_{ul}/g_{u}$ is the critical Hanle field in the collisionless regime 
(i.e., $B_H{\approx}23$ G for the Sr {\sc i} 4607 \AA\ line). We point out that in the atmospheric region of formation of the Sr {\sc i} 4607 \AA\ line (i.e., approximately between 250 km and 350 km for line of sights with $0.6\,{\ge}\,{\mu}\,{\ge}\,0.1$ in the MHD model) the horizontally averaged $B_c$ value varies between 60 G and 40 G, approximately, with ${\langle B_{c} \rangle}$ decreasing with height because ${\langle \delta^{(2)} \rangle}$ decreases with height in the solar atmosphere (following the density decrease). More or less similar $B_c$ values are found in one-dimensional semi-empirical models. It is also important to note that a {\em lower limit} for the Hanle saturation field of the Sr {\sc i} 4607 \AA\ line is $B_{\rm satur}=200\,{\rm G}\,{\approx}\,10\,B_H$, since full saturation is reached only for $B\,{\ge}\,300$ G. 

\section{Results}

The results reported here were obtained through the self-consistent numerical solution of the equations, applying the methods mentioned in Trujillo Bueno \& Shchukina (2007). In a first step we solved the scattering line polarization problem of the Sr {\sc i} 4607 \AA\ line in the MHD surface dynamo model, but imposing $B=0$ G at each spatial grid point. The resulting $Q/I$ and $U/I$ line-center signals are similar to those shown for three line-of-sights in figure 1 of Trujillo Bueno \& Shchukina (2007), which were obtained by solving the same 3D radiative transfer problem in the above-mentioned HD model. For the moment, the scattering polarization observations of the Sr {\sc i} 4607 \AA\ line that have been published lack spatial resolution (see Figure 2). We point out that when the $I$, $Q$ and $U$ profiles we have calculated in the MHD model are spatially averaged over scales significantly larger than that of the solar granulation pattern we obtain $U/I{\approx}0$ and the $Q/I$ line-center amplitudes given by the black open circles of Figure 2. Remarkably, these reference scattering polarization amplitudes are similar to those computed by Trujillo Bueno et al. (2004) in the HD model (see the green open circles).\footnote{For this reason we find again $\langle B \rangle\,{\approx}130$ G when the same hypotheses mention in \S1 are used to infer $ \langle B \rangle\ $ from calculations in the $B=0$ G version of the MHD model.} The small but noticeable differences between the two curves are due to the fact that the thermodynamic structure of the MHD model is not as accurate as in the HD model, with ${{\bar{J}^2_0}/{{\bar{J}^0_0}}}$ and $1/(1+\delta^{(2)})$ being slightly larger in the MHD model (in the photospheric region where the $Q/I$ profile of the strontium line originates, whose amplitude is approximately proportional to the product of such two quantities). 

The black dashed line shows the $Q/I$ amplitudes that we obtain when taking into account the Hanle depolarization produced by the actual magnetic field of the MHD model, whose magnetic Reynolds number is $R_{\rm m}{\approx}2600$. Not surprisingly, the Hanle depolarization produced by the model's magnetic field 
is too small to explain the observations. This is because the mean field strength at a height of about 300 km in the MHD model is only $\langle B \rangle\,{\approx}15$ G  (i.e., an order of magnitude smaller than the $ \langle B \rangle\ $ value inferred by Trujillo Bueno et al. 2004). Clearly, the level of small-scale magnetic activity of the MHD surface dynamo model is significantly weaker than that of the real quiet Sun photosphere. 

In order to estimate the mean field strength of the quiet Sun photosphere we scaled 
the magnetic strength at each spatial grid point by a constant factor, $F$, till fitting the observations of Figure 2. We have done this keeping the magnetic field topology of the MHD model fixed. We find $F{\approx}12$, which implies $\langle B \rangle\,{\approx}150$ G at a height of 300 km above the ``visible surface" in the MHD model (i.e., an average field strength similar to that inferred by Trujillo Bueno et al. 2004). Note that this result has now been reached without using the two hypotheses mentioned in \S1. 

Curiously enough, a similar scaling factor is also found 
when one follows Pietarila Graham et al. (2009) and assumes that the ratio between the magnetic strengths of the real solar photospheric plasma and of the original MHD surface dynamo model is equal to the square root of the ratio of the corresponding magnetic Reynolds numbers, with $R_{\rm m}({\rm Sun})\,{\approx}\,3{\times}10^5$ and $R_{\rm m}({\rm model})\,{\approx}\,2600$. As shown by the dotted line of Figure 3, 
to scale the field strength of the MHD model with a constant scaling factor $F{\approx}12$ implies an unrealistic height-variation of $ \langle B \rangle\ $. The ensuing large magnetic field strengths in the region of formation of the Zeeman signals of the Fe {\sc i} lines at about 15650 \AA\ and 6302 \AA\ (e.g., around 60 km in the MHD model for the Fe {\sc i} 6302.5 \AA\ line; see Figure 1) would produce synthetic Stokes profiles in contradiction with those observed. In fact, $F=12$ is significantly larger than the tentative scaling factor 3 needed by Danilovic et al. (2010) for explaining the histograms of the polarization signals produced by the Zeeman effect in the 6301.5 \AA\ and 6302.5 \AA\ Fe {\sc i} lines. On the other hand, $F=3$ is too low a value for explaining the scattering polarization observations of Figure 2 (see the dashed-dotted line). Therefore, in order to be able to fit the observational data of Figure 2 without destroying the possibility of explaining the Stokes profiles observed in the Fe {\sc i} lines, we have introduced a height-dependent scaling factor which leads to the solid line of Figure 3 for the height variation of the mean field strength. As shown by the black solid line of Figure 2, the magnetic field that results from our height-dependent scaling factor produces a Hanle depolarization in the Sr {\sc i} 4607 \AA\ line that explains well the observations\footnote{The small discrepancy between the black solid line and the data points of Figure 2 can be explained by the fact that the thermal and density structure of the MHD model is not as accurate as that of the HD model.}. It is also important to mention that a  significant fraction of the model's granular plasma that contributes to the scattering polarization of the C$_2$ lines mentioned in \S1 is magnetized with $ \langle B \rangle\ {\sim} 10\,{\rm G}$.

\section{Conclusions}

We have solved the radiative transfer problem of resonance polarization and the Hanle effect of the Sr {\sc i} 4607 \AA\ line in a 3D model of the quiet solar photosphere resulting from the magneto-convection simulations with surface dynamo action of V\"ogler \& Sch\"ussler (2007). We find that the level of magnetic activity in their surface dynamo model is too low for explaining the scattering polarization observations of the Sr {\sc i} 4607 \AA\ line. The observed linear polarization amplitudes can however be explained after multiplying each grid-point magnetic strength by a scaling factor $F{\approx}12$.

A scaling factor $F{\approx}12$ is about three or four times larger than that required   
to explain the polarization induced by the Zeeman effect in the Fe {\sc i} lines around 6302 \AA. This discrepancy between the scaling factors needed to explain the Hanle signals of the Sr {\sc i} 4607 \AA\ line and the Zeeman signals of the Fe {\sc i} lines disappears if we take into account that the Zeeman signals originate a few hundred kilometers deeper than the Hanle signals (see Figure 1). In other words, with a height-dependent scaling factor that implies $\langle B \rangle\,{\approx}\,160$ G in the low photosphere and $\langle B \rangle\,{\approx}\,130$ G in the upper photosphere it is possible to explain approximately both types of observations. 

Our investigation of the Hanle 
effect in the above-mentioned surface dynamo model 
reinforces the conclusions of Trujillo Bueno et al. (2004) summarized in \S1. 
In particular, the horizontal variation of the scaled magnetic field 
needed to fit the observations indicates that most of the flux
and magnetic energy reside on still unresolved scales in the intergranular plasma. 
Moreover, the fact that the magnetic energy density carried by this ``hidden" magnetic field is a significant fraction of the local kinetic energy density and that the scattering polarization amplitudes of the Sr {\sc i} line and of the 
C$_2$ lines do not seem to be clearly modulated by the solar cycle (Trujillo Bueno et al. 2004; Kleint et al. 2010), support the suggestion that a small-scale turbulent surface dynamo plays a significant role in the magnetism of the ``quiet" Sun.

\acknowledgments
{\bf Acknowledgments}
We are grateful to Manfred Sch\"ussler and Alexander V\"ogler 
for having kindly provided to us the surface dynamo model of the quiet solar photosphere we have used in this investigation. Finantial support by the Spanish Ministry of Science through projects AYA2010-18029 (Solar Magnetism and Astrophysical Spectropolarimetry) and CONSOLIDER INGENIO CSD2009-00038 (Molecular Astrophysics: The Herschel and Alma Era) is gratefully acknowledged.


\newpage


\begin{figure}
\plotone{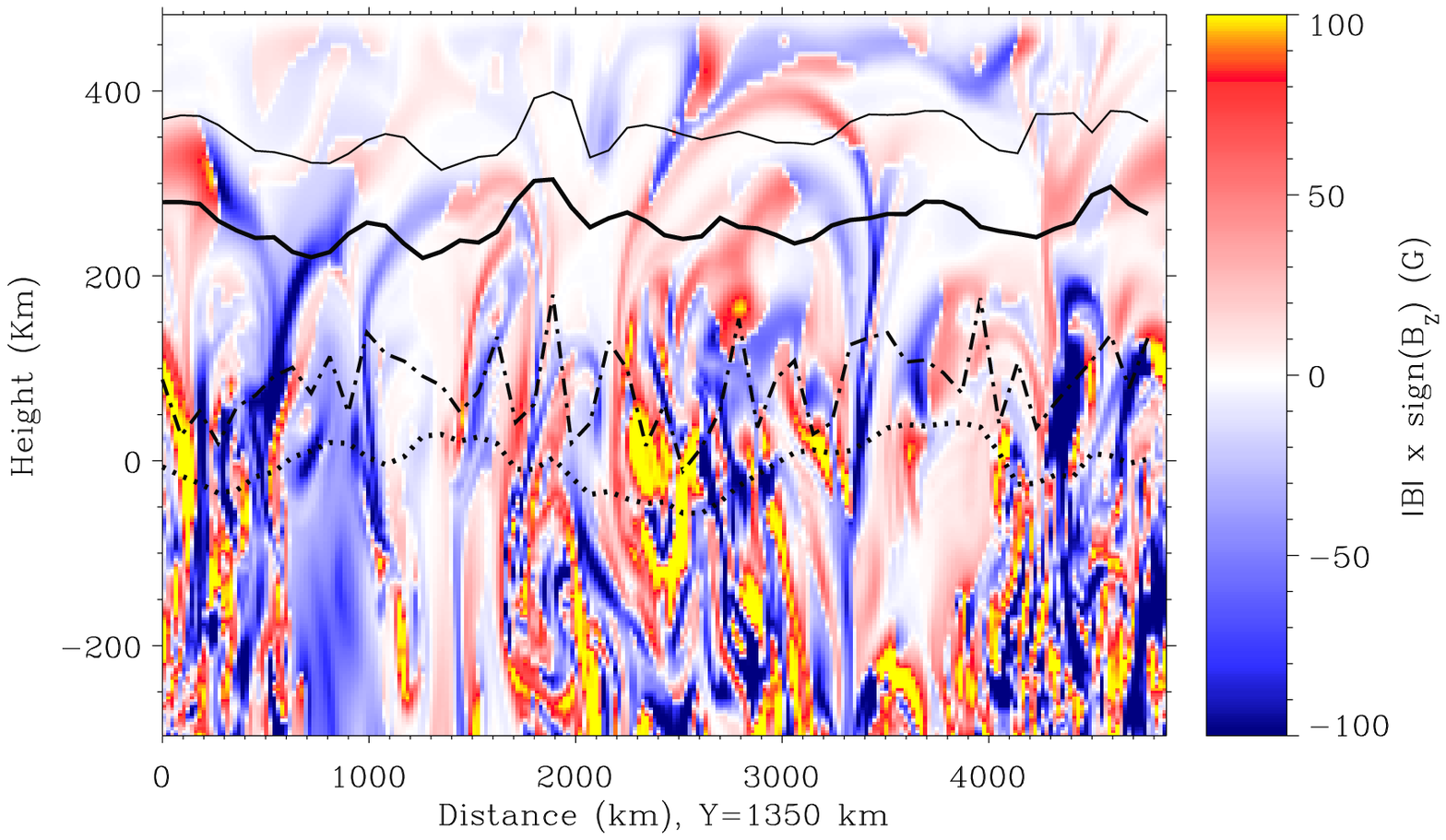}
\caption{The regions of formation, in the MHD model, of the (Stokes $Q$) Hanle signals of the Sr {\sc i} 4607 \AA\ line and of the (Stokes $V$) Zeeman signals of the Fe {\sc i} 6302.5 \AA\ line.
The colored background provides information on the vertical and horizontal variation of the 
model's magnetic strength, distinguishing between positive and negative values of the 
field's vertical component. The dotted line around $h{\approx}0$ km indicates the
height where the continuum optical depth at $\lambda{4607}$ is unity. The solid lines indicate the
heights where the line-center optical depth of the Sr {\sc i} 4607 \AA\ line is unity along line of sights with $\mu=0.6$ (thick line) and $\mu=0.1$ (thin line). The dashed-dotted line shows the heights where the optical depth of the Fe {\sc i} 6302.5 \AA\ line is unity along line of sights with $\mu=1$, calculated at the wavelengths where the modulus of the Stokes $V$ profiles have their maximum value. 
\label{fig:fig1}}
\end{figure}

\clearpage 

\begin{figure}
\plotone{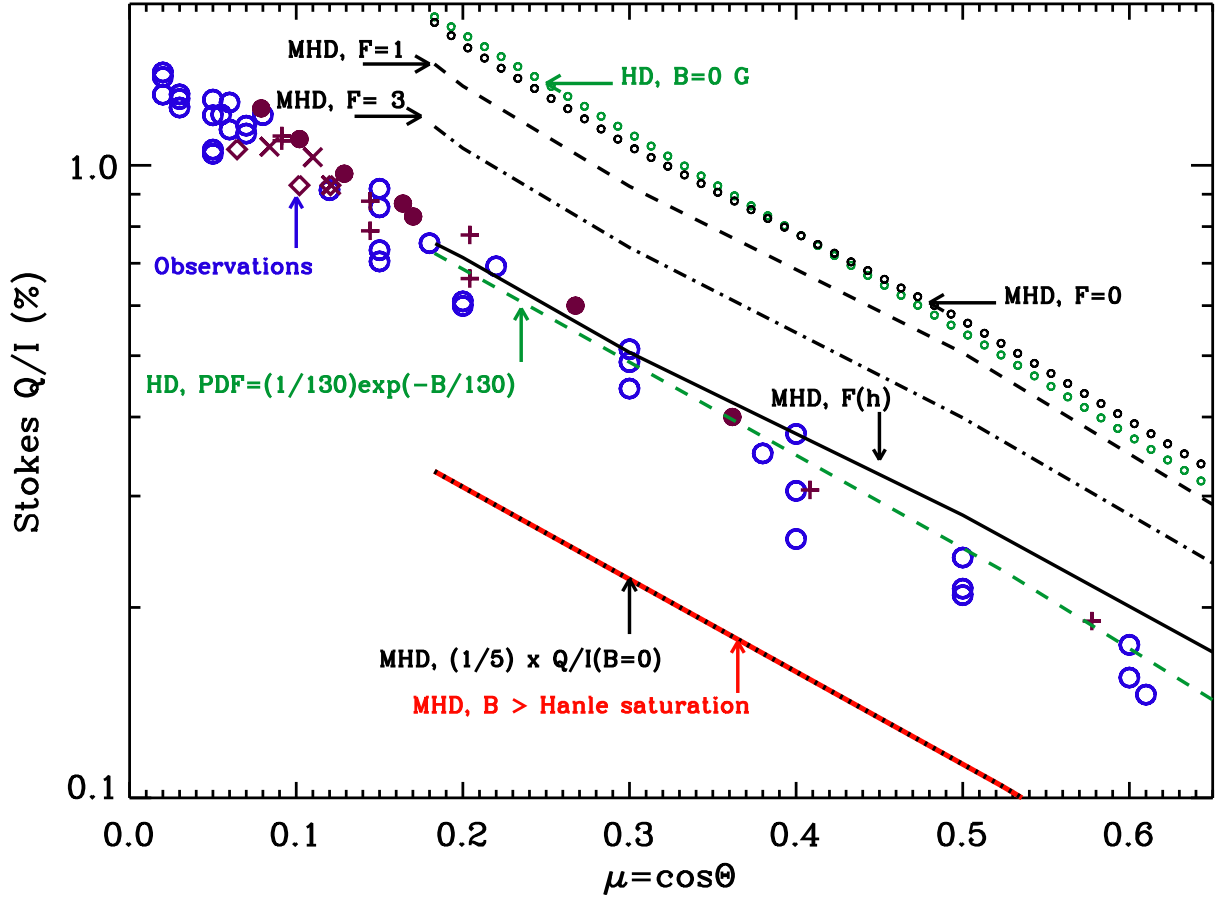}
\caption{Center-to-limb variation of the (spatially averaged)  
$Q/I$ scattering amplitudes of the photospheric line of Sr {\sc i} at 4607 \AA. 
The data correspond to various observations taken by several authors during a minimum and a maximum of the solar activity cycle (see references in Trujillo Bueno et al. 2004). The two green lines show scattering polarization amplitudes calculated in the HD model, without including any magnetic field (green open circles) and including the Hanle depolarization of a microturbulent field with an exponential PDF characterized by a mean field strength $\langle B \rangle {=} 130$\,G (green dashed line). The black lines show scattering polarization amplitudes calculated in the MHD model, neglecting its magnetic field (black open circles) and taking into account the Hanle depolarization of the model's magnetic field with scaling factors $F=1$ (black dashed line) and $F=3$ (black dashed-dotted line). As shown by the black solid line, the observations can be approximately fitted by multiplying each grid-point magnetic strength by a height-dependent factor $F(h)$ (which implies the height-variation of $\langle B \rangle$ given by the solid line of Figure 3). Practically the same black solid line is obtained with a constant scaling factor $F=12$ (which implies the unrealistic height-variation of $\langle B \rangle$ given by the dotted line of Figure 3). The red solid line shows the calculated scattering polarization amplitudes when imposing $B>B_{\rm satur}=300$ G at each grid point in the MHD model. The fact that this red line coincides with 1/5 of the zero-field $Q/I$ amplitudes indicates that the microturbulent field approximation is indeed a suitable one.
\label{fig:fig2}}
\end{figure}

\clearpage

\begin{figure}
\plotone{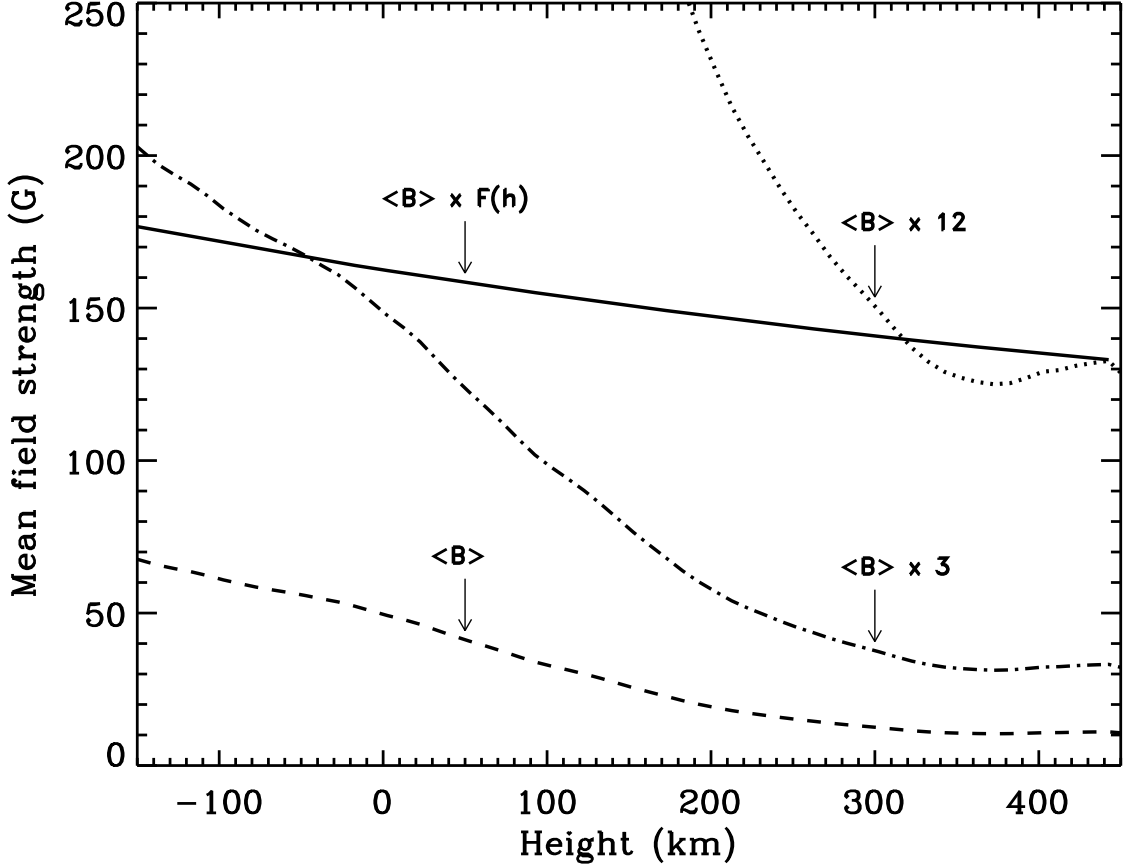}
\caption{Various height-variations of the mean field strength corresponding to four different scaling factors $F$ of the magnetic strength of the MHD model. Dashed-line: $F=1$ (i.e., as in the original MHD model). Dashed-dotted line: $F=3$ (i.e., as suggested by the Zeeman polarization investigation of Danilovic et al. 2010). Dotted line: $F=12$. Solid line: a height-dependent scaling factor that explains the scattering polarization observations of the Sr {\sc i} 4607 \AA\ line without destroying the possibility of explaining simultaneously the polarization induced by the Zeeman effect in the Fe {\sc i} lines. The arrows at $h{\approx}60$ km and $h{\approx}300$ km indicate the approximate atmospheric heights around which the observed Zeeman and Hanle signals are produced, respectively. Note from the solid and dashed lines that $F(h=60\,{\rm km})\,{\approx}\,4$ while $F(h=300\,{\rm km})\,{\approx}\,10$.
\label{fig:fig3}}
\end{figure}

\clearpage

\end{document}